\pdfoutput=1

\documentclass[11pt]{article}

\usepackage[final]{acl}

\usepackage{times}
\usepackage{latexsym}

\usepackage[T1]{fontenc}

\usepackage[utf8]{inputenc}

\usepackage{microtype}

\usepackage{inconsolata}

\usepackage{graphicx}
\usepackage{pgfplots}
\usetikzlibrary{patterns}
\pgfplotsset{compat=1.18}
\usepackage{multirow}
\usepackage{subcaption}
\usepackage{CJKutf8}
\usepgfplotslibrary{statistics}
\usepackage[most]{tcolorbox}
\title{Continual Pretraining on Encrypted Synthetic Data for Privacy-Preserving LLMs}

\author{
 \textbf{Honghao Liu\textsuperscript{1,2}},
 \textbf{Xuhui Jiang\textsuperscript{1,4}},
 \textbf{Chengjin Xu\textsuperscript{1,4}},
 \textbf{Cehao Yang\textsuperscript{1,2}},
 \textbf{Yiran Cheng\textsuperscript{3}},
 \\
 \textbf{Lionel Ni\textsuperscript{2,3,}\thanks{Corresponding authors.}},
 \textbf{Jian Guo\textsuperscript{1,2,}\footnotemark[1]}
\\
\\
 \textsuperscript{1}International Digital Economy Academy\\
 \textsuperscript{2}The Hong Kong University of Science and Technology (Guangzhou)\\
 \textsuperscript{3}The Hong Kong University of Science and Technology\\
 \textsuperscript{4}DataArc Tech Ltd.
}

\begin{document}
\maketitle
\begin{abstract}
Preserving privacy in sensitive data while pretraining large language models on small, domain-specific corpora presents a significant challenge. In this work, we take an exploratory step toward privacy-preserving continual pretraining by proposing an entity-based framework that synthesizes encrypted training data to protect personally identifiable information (PII). Our approach constructs a weighted entity graph to guide data synthesis and applies deterministic encryption to PII entities, enabling LLMs to encode new knowledge through continual pretraining while granting authorized access to sensitive data through decryption keys. Our results on limited-scale datasets demonstrate that our pretrained models outperform base models and ensure PII security, while exhibiting a modest performance gap compared to models trained on unencrypted synthetic data. We further show that increasing the number of entities and leveraging graph-based synthesis improves model performance, and that encrypted models retain instruction-following capabilities with long retrieved contexts. We discuss the security implications and limitations of deterministic encryption, positioning this work as an initial investigation into the design space of encrypted data pretraining for privacy-preserving LLMs. Our code is available at  \href{https://github.com/DataArcTech/SoE}{https://github.com/DataArcTech/SoE}.
\end{abstract}

\section{Introduction}
Large language models (LLMs) have demonstrated remarkable success across a wide range of natural language processing tasks, largely due to their ability to learn from vast and diverse datasets \cite{openai2024gpt4omini, gemini2024google, llama32024Grattafiori, liu2024deepseek}. Despite these advances, two significant challenges persist in their real-world deployment: learning efficiently from rare or niche information such as small private corpora and preserving data privacy \cite{pmlr-v202-kandpal23a,rho2025encryptionfriendly,Sebastian2023privacy}. As LLMs are continuously training on new, small-scale, and potentially sensitive data, they may struggle to sufficiently acquire new knowledge or risk forgetting previously learned information. Furthermore, training LLMs on sensitive data containing personally identifiable information (PII) introduces substantial privacy risks \cite{Sebastian2023privacy, Peris2023privacy}.

\begin{figure}
    \centering
    \includegraphics[width=1\linewidth]{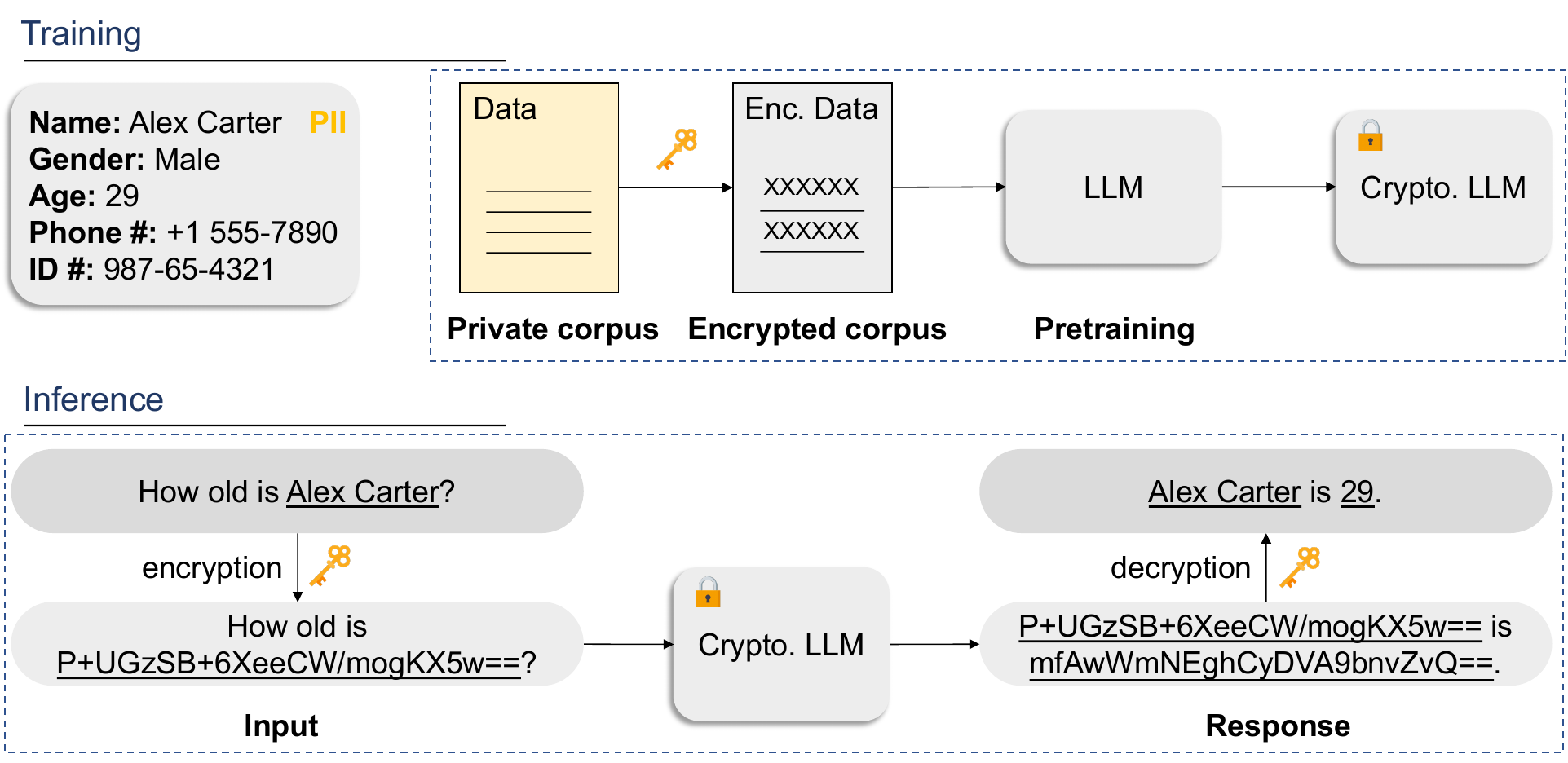}
    \caption[Caption for LOF]{The task for privacy-preserving LLMs\footnotemark.}
    \label{fig: task}
\end{figure}
\footnotetext{The PII shown in the figure is synthetic and not real.}
A promising solution to address the challenge of limited data is the use of synthetic data to pretrain LLMs \cite{yang2024entigraph}. High-quality synthetic data can enhance the efficiency and effectiveness of LLM pretraining, allowing the model to learn useful patterns and knowledge even from small-scale data. However, generating synthetic data that simultaneously preserves semantic richness and protects sensitive information remains challenging. Cryptographic approaches such as homomorphic encryption enable computation over encrypted embeddings, model weights, or gradients, but are typically applied to inference or fine-tuning scenarios and incur substantial computational overhead \cite{luo2020glyph,hou2023ciphergpt,liu2024llencprompt,rho2025encryptionfriendly}. Other mainstream privacy-preserving strategies -- such as data curation, contextual prompting, post-training alignment, and differential privacy -- can mitigate leakage risks but do not support controlled recovery of sensitive information by authorized users through cryptographic keys \cite{Lukas2023,xiao-etal-2024-large,asthana2025adaptive,li2025privacyft}.

Motivated by these limitations, we explore the feasibility of training LLMs directly on encrypted data, where the cloud provider cannot access the plaintext of private corpora. As illustrated in Figure \ref{fig: task}, LLMs are continually pretrained on encrypted data to memorize ciphertext representations while learning the underlying domain knowledge embedded in the content. During inference, user queries are similarly encrypted and processed by the pretrained LLM (Crypto. LLM) to generate the encrypted answers. The user with private key can then decrypt the ciphertext to get the response. In this framework, the cloud server remains unaware of the private information, ensuring that sensitive information is protected from exposure.

\begin{figure*}[htbp]
  \includegraphics[width=0.96\linewidth]{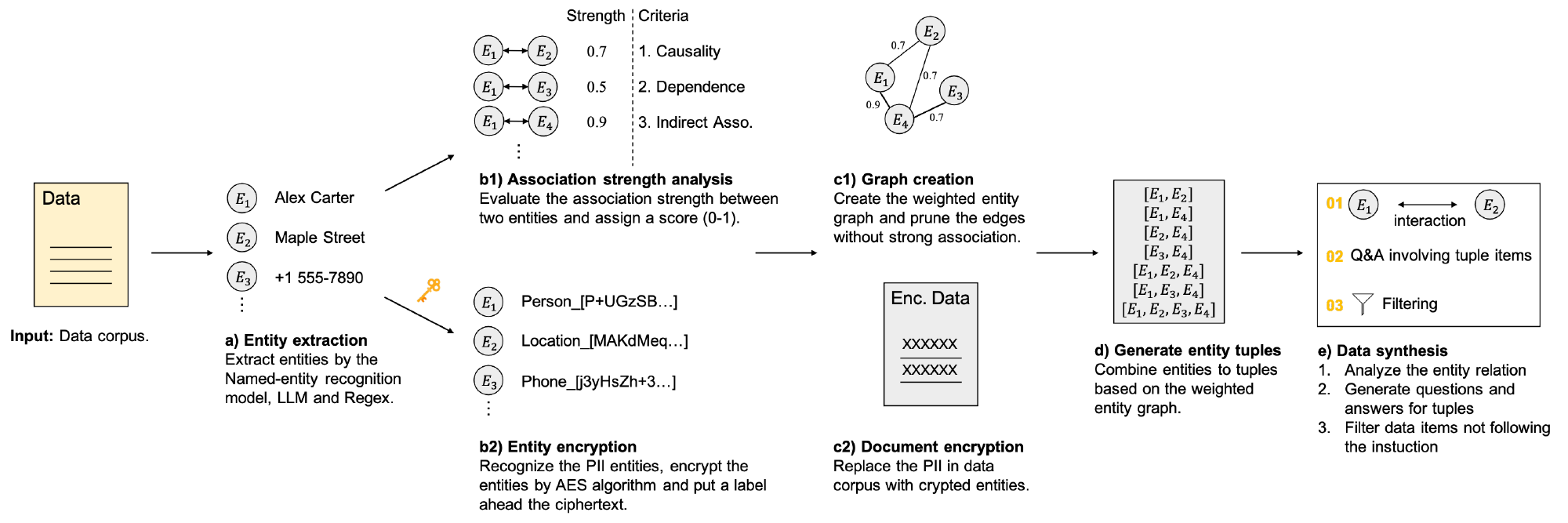}
  \caption {A privacy-preserving approach for LLM pretraining by synthesizing data with entities, encrypting PII using the AES, and training the model on the encrypted data.}
  \label{fig: workflow}
\end{figure*}

In this paper, we propose an exploratory  entity-based data synthesis framework for privacy-preserving continual pretraining of LLMs. Our approach involves several key steps to ensure privacy while enabling continual learning. First, we extract entities using a combination of GPT and named entity recognition (NER) models \cite{Zhao2019ner,Xu2020ner}. We then construct a weighted entity graph to capture entity associations and generate strongly related entity tuples. Using large language models such as GPT and DeepSeek, we synthesize question–answer pairs and relational data conditioned on these tuples, followed by filtering to remove noncompliant samples \cite{openai2024gpt4omini,2024deepseekv2, liu2024deepseek}. We encrypt PII entities using AES encryption to protect sensitive information \cite{daemen1999aes}. Since the encryption could perform before synthesis, we manually verify the PII encryption on the small-scale original data to minimize unintended privacy leakage. Figure \ref{fig: workflow} provides a concrete illustration. Finally, we pretrain  Llama3-8B and Qwen2.5-7B on the encrypted synthetic data, enabling the model to interact with with ciphertext representations and support controlled decryption by authorized users \cite{llama32024Grattafiori, qwen2025qwen25technicalreport}.

We evaluate the effectiveness of our method through four sets of experiments on QuALITY - a general English dataset and a Chinese Judge case dataset with private QA pairs \cite{Pang2022quality,judge2024spc}. First, we examine how entity-related factors, including the number of entities and the length of entity relation tuples, influence downstream question answering performance. We find that the number of entities correlates with performance, while the relation length also has a notable effect. This demonstrates the effectiveness of our weighted-graph-based synthetic method, which outperforms the EntiGraph \cite{yang2024entigraph}. Second, we show that the crypto models evaluated on encrypted questions significantly outperform based models, with the performance drops substantially on unencrypted questions, confirming that PII is securely preserved (see Section \ref{sec: experiment_pii_preserve}). Third, we compare base models, models trained on unencrypted synthetic data, and models trained on encrypted synthetic data (crypto models). Although encrypted models incur a modest performance drop relative to unencrypted synthetic models, they consistently outperform base models, demonstrating the feasibility of privacy-preserving continual pretraining. Finally, under a retrieval-augmented generation (RAG) setting \cite{lewis2020rag, gao2023rag}, encrypted models retrieve encrypted data efficiently, achieving performance comparable to base models and, in some cases, exceeding that of unencrypted models.

The  contributions of this paper are as follows.
\begin{itemize}
    \item We propose an entity-based, weighted-graph data synthesis framework combined with deterministic encryption as an exploratory approach to privacy-preserving continual pretraining of LLMs, enabling controlled access to sensitive data via cryptographic keys. 
    \item We demonstrate, on two datasets including the Quality dataset and a Chinese Judge case dataset with private QA, that LLMs can acquire new knowledge through continual pretraining on encrypted synthetic data.
    \item We conduct a series of experiments to analyze the impact of entity characteristics, the effectiveness of encryption in preserving PII, and the ability on retrieval settings.
    \item We demonstrate that while models trained on encrypted data exhibit a modest performance trade-off compared to unencrypted synthetic training, they substantially outperform base models while protecting sensitive information.
\end{itemize}

Overall, this work positions encrypted synthetic data pretraining as a promising yet underexplored direction for privacy-aware LLM adaptation. We emphasize that our study represents an initial investigation into the design space. 
\section{Related work}
\subsection{Pretraining data synthesis}
Recent works have explored various synthetic data generation techniques to enhance LLM pretraining efficiency and effectiveness. \citet{Eldan2023tiny} introduced TinyStories, a dataset of short synthetic stories, demonstrating that models trained on it can generate fluent and diverse narratives while retaining reasoning capabilities. The Phi model series are prposed to leverage  synthetic data for pretraining, progressively enhancing performance in coding, reasoning, and multimodal tasks \citep{Gunasekar2023textbooks,Li2023phi1.5,abdin2024phi3}. \citet{Maini2024rephrase} further improved synthetic data quality by rephrasing web documents in diverse styles, ensuring that pretraining does not rely solely on LLMs for information. Recently, \citet{yang2024entigraph} introduced EntiGraph, a synthetic data augmentation algorithm that synthesizes effective training data based on the underlying knowledge within entity graphs.

Beyond model pretraining, researchers have explored synthetic data techniques for instruction tuning and augmentation. \citet{li2024self} and \citet{wei2024magicoder} leveraged synthetic data for instruction tuning and coding enhancement. \citet{Jung2024distill} proposed the Impossible Distillation framework, which refines low-quality teacher outputs into high-quality paraphrases and summarizations by reducing the LLM search space. 
\subsection{Privacy Preserving in LLMs}
There are three primary approaches to protecting private information in LLMs: (1) curating the pretraining corpus to remove sensitive data \cite{asthana2025adaptive,Lukas2023}, (2) controlling LLMs by differential privacy (DP) or contextual prompting \cite{yu2024privacyalign,zhang2024enhancing,kim2023propile}, and (3) aligning models through reinforcement learning from human feedback (RLHF) or fine-tuning  \cite{li2025privacyft, xiao-etal-2024-large}. However, DP-based training, contextual prompting, and alignment methods may still result in privacy leakage, as models typically have access to private information during the pretraining. Furthermore, none of these methods supports authorized users in retrieving private responses while restricting access for unauthorized users.

Homomorphic encryption (HE) provides a feasible solution to encrypt token embeddings, weights and gradients to enable ciphertext computation, but it normally is typically used for model inference or finetuning and faces challenges such as significant computational overhead \cite{rivest1978data, bachrach16cryptonets, luo2020glyph}. Recent work accelerates the HE LLM by model architecture modification, communication optimization and computation approximations \cite{rho2025encryptionfriendly,liu2024llencprompt,hou2023ciphergpt}. Nonetheless, performing matrix multiplications on encrypted tensors remains computationally intensive for LLM pretraining, and it requires LLMs to have a polynomial form \cite{vats2023privacy,Peris2023privacy}. In 2022, CipherDAug uses ciphertext for data augmentation that improves neural machine translation \citep{kambhatla2022cipherdaug}. \citet{christodoulou2024med} trained the MED with encrypted data to complete encrypted sentences based on a short prompt; however, it strictly memorizes encrypted whole sentences and cannot reason or extract underlying knowledge. 

In contrast, we propose a novel framework that encrypts PII and trains LLMs to memorize and learn internal knowledge in its encrypted form. This approach enables authorized users with the appropriate decryption keys to access private outputs, while preventing unauthorized access. Moreover, it avoids the significant computational overhead associated with homomorphic encryption.


\section{Method}

This study aims to address continual pretraining from a small corpus and the privacy challenges in LLMs through efficient, high-quality data synthesis. Figure~\ref{fig: task} illustrates the problem setting.
We first describe the overall workflow of our approach (\ref{sec: method_workflow}). Then, we present the PII entity extraction and encryption and decryption method in detail (\ref{sec: method_crypto}). Next, we describe the entity-based data synthesis process and discuss privacy guarantee and leakage in this process (\ref{sec: method_synthesis}). 

\subsection{Workflow}
\label{sec: method_workflow}
The workflow of our approach consists of several key stages aimed at preserving privacy while enabling continual learning in large language models, see Figure~\ref{fig: workflow}. First, entities are extracted from the data corpus using the Named Entity Recognition (NER) model, regular expression techniques and LLMs. Then we recognize the PII entities, such as names, locations, phone numbers, and other sensitive information. These entities will be encrypted using AES encryption with a private key on synthetic data, ensuring that sensitive information remains secure. We analyze the association strength between entities and create a weighted graph. Following graph creation, synthetic data generation on entity tuples is carried out by large language models, such as GPT 4o and DeepSeek, which generate questions and answers, analyze entity relations, and filter irrelevant data. The resulting synthetic encrypted data is used to pretrain LLMs.

\subsection{PII preservation}
\label{sec: method_crypto}
Preserving the privacy of personally identifiable information (PII) is essential when training models on sensitive data. In this work, we employ a combination of entity extraction and encryption techniques to ensure that PII is securely handled and protected throughout the process.

\textbf{PII extraction}. To identify PII entities, we use a combination of named entity recognition (NER) models \citep{honnibal2017spacy,MsPresidio,Zhao2019ner,Xu2020ner} and regular expression patterns. NER models are employed to recognize entities such as names, locations, and other contextual PII, while regex patterns help identify more structured PII, including phone numbers and bank card numbers. These extraction techniques are integrated into the Microsoft Presidio toolkit \citep{MsPresidio}, which facilitates the  extraction of sensitive information from text. 

\textbf{PII encryption}. To ensure the privacy of PII, we apply the symmetric encryption algorithm AES in ECB mode to encrypt the extracted PII entities \citep{daemen1999aes,dworkin2001recommendation,rijmen2001advanced}. We choose ECB mode intentionally because it produces deterministic ciphertext - the same plaintext entity is always encrypted to the same ciphertext. This consistency is crucial for allowing the LLM to learn entity-level associations and maintain coherence. In contrast, more secure modes like CBC introduce randomness, resulting in non-deterministic ciphertexts. This makes it impossible for the model to recognize that multiple different-looking ciphers refer to the same underlying entity. PII in both synthetic documents and the original corpus is encrypted using this algorithm, ensuring that sensitive data remains secure. After encryption, the ciphertext is encoded in Base64 format \citep{wiki:Base64}, which includes a mix of alphabets, numbers, and special characters. To further improve the integrity and interpretability of the encrypted data, we prepend a prefix to the encrypted PII. For example, names are encrypted and formatted as \texttt{Person\_[cipher\_name]}, ensuring that the PII remains secure and distinguishable.
\subsection{Entity-based synthetic data generation}
\label{sec: method_synthesis}
In generating entity-based synthetic data, we follow the procedures outlined in Entigraph \citep{yang2024entigraph}, with several key modifications designed to enable to more efficient and comprehensive data generation. The process consists of three main steps: entity extraction, entity description, and relation analysis. Below, we describe the adaptations we made to the standard Entigraph approach:

\textbf{Entity extraction}. In contrast to Entigraph, which relies solely on LLMs for entity extraction, we utilize a combination of LLMs, along with NER models. Our experiments show that extracting more comprehensive entities improves model performance, especially as the number of entities increases during continual pretraining. Relying only on LLMs for extraction may miss key entities, limiting the data coverage and model performance.

\textbf{Relation analysis}. While Entigraph focuses on generating pairwise and 3-tuple relationships due to the computational intensity of considering all possible tuples, we extend this approach by generating a wider range of k-tuples ($k<N$ and $N$ is the total number of entities). We use LLMs to generate an association strength score which contains causality, dependence and indirect association as criteria for each pair of entities, creating a weighted graph of relations. A threshold is applied to filter out weak edges, leaving only the strongest relations. k-tuples are created by selecting its $k-1$ most strongly associated neighbors. In Appendix \ref{appendix_prompt}, we provide the prompt used to obtain the association strength scores. This allows us to generate more comprehensive relations among entities. 

\textbf{Data filtering}. After generating the synthetic data, we apply filtering techniques to eliminate samples that do not adhere to the instruction. This step ensures that the synthesized data is high-quality and relevant to the task at hand, thereby improving the effectiveness of the pretraining process \cite{li2024datas_election, xie2023data_selection, Tirumala2023data_selection}.

\textbf{Privacy leakage and privacy considerations.}
If entity recognition and encryption are applied after data synthesis, privacy protection depends heavily on the recall of named entity recognition (NER) models. In such cases, high-recall NER models can detect most names and locations, while regular expressions capture numerical identifiers such as dates and identification numbers. Nevertheless, rare, ambiguous, or context-dependent entities may still be missed, resulting in residual PII and potential privacy leakage (see Appendix~\ref{appendix_privacy_leakage} for empirical statistics).

To mitigate this risk, we could perform PII detection and encryption \emph{prior} to data synthesis. The original corpora considered in this study are small and manageable, allowing us to apply a combination of LLM-based and NER-based entity extraction followed by manual verification. All identified PII entities are encrypted before any synthetic data generation is performed. Consequently, data synthesis is conditioned solely on encrypted representations, substantially reducing the exposure of sensitive information during generation and subsequent training. While this design does not claim formal cryptographic security guarantees, it provides a practical and controlled privacy protection mechanism suitable for exploring the feasibility of continual pretraining on encrypted data. We note that our use of deterministic encryption is motivated by functional requirements for consistency during continual pretraining; the security trade-offs and limitations of this choice are discussed in the Limitations section \ref{sec:limitation}.

By incorporating these modifications, we ensure that the synthetic data generation process is comprehensive, efficient and supporting continual learning with preserving privacy.



\section{Experiments \& Evaluation}
In this section, we describe the experimental setup (\ref{sec: experiment_setup}) and conduct a comprehensive evaluation of our proposed approach. Specifically, we assess the impact of entity-related factors on model performance (\ref{sec: experiment_factors}), the effectiveness of our method in preserving personally identifiable information (PII) (\ref{sec: experiment_pii_preserve}), the comparison of different model performances (\ref{sec: experiment_perf_comp}), and the assessment of retrieval-augmented generation (RAG) capabilities (\ref{sec: experiment_rag}).
\subsection{Experimental setup}
\label{sec: experiment_setup}

\textbf{Data corpus}. The data corpus used for experiments consists of two primary components, a subset of the Quality dataset and the Chinese Judge case dataset \citep{Pang2022quality, judge2024spc}. Quality is a general purpose dataset with multiple-choice QA in English, and we sort the Chinese Judge case dataset from \citep{judge2024spc} as the primary corpus to evaluate model performance. We designed the QA pairs for the Chinese Judge dataset to assess the models' ability to handle sensitive real-world data. Details of the Chinese Judge case dataset are described in the Appendix \ref{appendix_chinese_dataset}.

We evaluate our results on English articles with 125 questions and Chinese articles with 72 questions in the Quality and Judge datasets. We synthesize data for continual pretraining with 10M to 30M tokens generated from an original source text of 65K tokens. The reason for using a small number of tokens is that Entigraph shows log-linear scaling of accuracy with all articles \cite{yang2024entigraph}. This setup allows us to analyze the effectiveness of continual pretraining in domain-limited scenarios, which are relevant in privacy-preserving and cost-sensitive applications.


\textbf{Synthesis and training}. We first synthesize the data by extracting entities using LLMs and NER models \cite{honnibal2017spacy,Xu2020ner,Zhao2019ner} (the en\_core\_web\_lg spaCy model for English NER), generating relevant question-answer pairs and relationships using GPT-4o mini \citep{openai2024gpt4omini} and Deepseek \citep{2024deepseekv2} for English and Chinese data generation. The data is synthesized from the original corpus to continually pretrain LLama3 8B and Qwen2.5 7B \cite{llama32024Grattafiori, qwen2025qwen25technicalreport}. We synthesize the data once for each model training, as it is time-consuming and expensive to synthesize by calling LLM APIs. We continually pretrain models on 8 Nvidia A100 GPUs, each with 40GB device memory with parameter settings in Appendix \ref{appendix:train_parameter}. We do not evaluate additional model sizes for two main reasons. First, in our initial experiments using LLaMA3 2B with only names encrypted \cite{llama32024Grattafiori}, the model’s performance was only marginally better than random guessing. Second, due to limited computational resources, we did not extend pretraining to larger models such as the 13B or 70B variants.
\subsection{The effect of entity-related factors}
\label{sec: experiment_factors}
We explore how various entity-related factors influence the performance of the model during continual pretraining. Specifically, we evaluate the number of entities, and the length of relation tuples to determine their impact on the quality of the synthetic data and model performance.

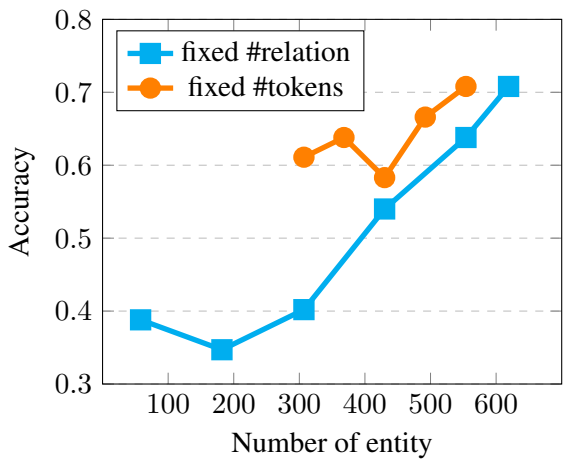
\begin{figure}[ht]
\centering
\begin{tikzpicture}
\begin{axis}[
    xlabel={Number of entity},
    ylabel={Accuracy},
    width=\columnwidth*0.99,
    height=\columnwidth/1.2,
    ymin=0.3, ymax=0.8,
    ytick distance=0.1,
    xmin=0, xmax=700,
    xtick={100,200,300,400,500,600},
    legend pos=north west,
    legend style={nodes={scale=1, transform shape}},
    ymajorgrids=true,
    grid style=dashed,
]
\addplot [color=cyan,mark=square*, line width=2pt,mark size=3pt] coordinates {
    (58,0.388) (182,0.347) (307,0.402) (430,0.54) (554,0.638) (619,0.708)
};
\addlegendentry{fixed \#relation}
\addplot [color=orange,mark=*,line width=2pt,mark size=3pt] coordinates {
  (307,0.611) (368,0.638) (430,0.583) (492,0.666) (554,0.708)
};
\addlegendentry{fixed \#tokens}
\end{axis}
\end{tikzpicture}
\caption{The effect on number of entities.}
\label{fig: experiment_num_entities}
\end{figure}
\textbf{Number of entities}. We investigate how the number of entities in the synthetic data affects the model's performance. To do this, we vary the number of entities used for data synthesis and evaluate the model's performance on a QA task. Our hypothesis is that a higher number of entities will lead to improved performance, as the synthetic data comprehensively covers the knowledge in the original data corpus. Figure \ref{fig: experiment_num_entities} illustrates the effect of the entity number on the Chinese Judge case QA. We vary the entity number while keeping the number of relation constant (blue squares). The fixed \#tokens refers to the fixed number of tokens at 20M in synthetic data (orange circles). Both curves show a general increasing trend in accuracy, but the fixed \#relation curve demonstrates a smoother and more consistent increase compared to the fluctuations in the fixed \#tokens curve.

\textbf{Length of tuples}.
Next, we examine the effect of the relation tuple length on performance. Relation tuples contain a certain number of entities and represent connections between them including relationships, interactions and dependencies. We vary the maximum length of these tuples and generate tuples according to the weighted graph. To assess the impact of tuple length, we evaluate the  performance of models on a QA task.
\begin{table}[hbpt]
    \centering
    \begin{tabular}{cccccc}
    \hline    
        Tuple length & 4&6&8&10 \\
    \hline
        Org. acc. & 0.67 & 0.69 & 0.65 & 0.72 \\
        Enc. acc. & 0.54 & 0.61 & 0.60 & 0.68 \\
    \hline
    \end{tabular}
    \caption{The effect on the length of entity tuple.}
    \label{tab: tuple_length}
\end{table}
We prune edges between two entities if their association strength is less than 0.5. We maintain approximately 22K generated relations while varying the maximum tuple length. Our results show that there is a performance increase as the length of relation tuples reaches 10 in both original and encrypted synthetic data pretraining. Prior to this point, there is a fluctuation as the length increases. This trend is illustrated in Table \ref{tab: tuple_length}. The highest accuracy is achieved when the tuple length is set to 10, with scores of 0.72 and 0.68, which surpass the results obtained using EntiGraph generation (0.68 and 0.61).


\subsection{PII preservation}
\label{sec: experiment_pii_preserve}
In this subsection, we investigate whether encrypted data preserves PII.

\textbf{Judge case dataset results}. We designed multiple-choice questions related to privacy within the dataset.
Then, we performed an experiment to test whether the PII entities are critical for answering the questions we designed by assessing whether the model requires access to PII entities to answer the questions correctly. The experimental setup involves testing the model's ability to answer these PII-related questions when the PII entities are masked, thereby simulating a scenario in which the model does not have access to private information.

\begin{table}[htbp]
    \centering
    \begin{tabular}{cccccc}
    \hline
        Model & Pretrain & Mask & Synt. & Accu. \\
    \hline
        & T & F & T & 0.68 \\
        Qwen2.5 & T & T & T & 0.263 \\
        7B & T & T & F & 0.208 \\
         & F & T & N/A & 0.208 \\
    \hline
    \end{tabular}
    \caption{The importance of PII for question answering in the Judge case dataset.}
    \label{tab:experiment_mask_PII}
\end{table}
The results, shown in Table \ref{tab:experiment_mask_PII}, demonstrate that when the PII entities are masked, the base model performs with an accuracy of only 0.208, which is the same as the model pretrained with the original data. The responses of synthetic model on masked questions show a slightly higher performance than the base model (close to random selection - 0.25), but exhibit a significant drop compared to the synthetic model with unmasked questions. This confirms that our PII-related questions are indeed private and require the PII  to answer accurately.
\begin{table*}[thb]
    \centering
    \begin{tabular}{cccccc}
    \hline
        Model & CPT & Synthesis & Data Encryption & Question Encryption & Accuracy \\
    \hline
         & T & T & T & T & 0.592 \\
         & T & T & F & F & 0.63 \\
         Llama3 8B& F & N/A & N/A & F & 0.424 \\
         & T & F & T & T & 0.424 \\
         & T & F & F & F & 0.488 \\
    \cline{2-6}
         \multirow{ 6}{*}{Qwen2.5 7B} & T & T & T & T & 0.611 \\
         & T & T & T (All) & T & 0.569 \\
         & T & T & F & F & 0.68 \\
         & F & N/A & N/A & F & 0.388 \\
         & T & F & T & T &  0.402 \\
         & T & F & F & F & 0.33 \\
    \hline
    \end{tabular}
    \caption{Performance comparison among models on Quality and Judge case datasets.}
\label{tab:experiment_performance_comp}
\end{table*}

Next, we compare the performance of three models:
the base model, the crypto model pretrained with encrypted synthetic data with encrypted questions (encS.) and the crypto model pretrained with encrypted synthetic data but evaluated with plaintext questions (noEncQ.). Our assumption is that if the crypto model can only answer encrypted questions, but not their unencrypted counterparts, then it is unaware of the private information.

\begin{figure}[tbph]
\centering
    \begin{tikzpicture}
    \begin{axis}[ybar,
        bar width=0.5cm, 
        width=\columnwidth,
        height=\columnwidth/1.5,
        ylabel = {Accuracy},
        ymin=0.25, ymax=0.9,
        ytick={0.3,0.4,0.5,0.6,0.7},
        xmin=0.5, xmax=3.7,
        xtick = {1,2,3.2},
        xticklabels={
        Quality,
        Judge Case,
        PII Quality
        },
        major x tick style = transparent,
        legend pos=north west,
        legend columns=4,
        legend style={nodes={scale=0.855, transform shape}},
        ymajorgrids=true,
        grid style=dashed,
        nodes near coords,
        every node near coord/.append style={font=\tiny},
        point meta=rawy,
        every node near coord/.append style={inner xsep=1pt,yshift=-2pt, /pgf/number format/.cd,fixed zerofill,precision=2},
    ]
    
    \addplot [color=black,postaction={pattern=north east lines}]  plot[ error bars/.cd, y dir=plus, y explicit] 
    coordinates {(1,0.42) (2,0.38) (3.2,0.411)};
    \addlegendentry{base}
    \addplot [color=black,fill=cyan]  plot[ error bars/.cd, y dir=plus, y explicit] 
    coordinates {(1,0.60) (2,0.569) (2,0.611) (3.2,0.529)};
    \addlegendentry{encS.}
    \addplot [color=black,fill=brown!30]  plot[ error bars/.cd, y dir=plus, y explicit] 
    coordinates {(1,0.584) (2,0.375) (3.2,0.441)};
    \addlegendentry{noEncQ.}
    \draw[black,dashed] (axis cs:2.6,0.25) -- (axis cs:2.6,0.9);
    \end{axis}
    \end{tikzpicture}
\caption{Performance comparison between base and crypto models.}
\label{fig: experiment_PII_preserve}
\end{figure}
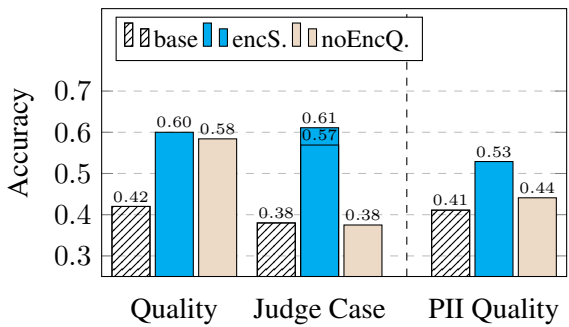
Figure \ref{fig: experiment_PII_preserve} illustrates the performance of these models in terms of PII preservation. The encS. model significantly outperforms the base model. In contrast, the noEncQ. model evaluated with plaintext questions shows no performance improvement compared to the base model. These results indicate that the encrypted model does not leak additional information and is unable to recognize the relationship between the plaintext and its ciphertext, thereby preserving PII.

\textbf{Quality dataset results}. Since the Quality dataset contains little sensitive information, we briefly analyze its performance to validate our privacy preservation findings. As illustrated in Figure \ref{fig: experiment_PII_preserve}, the encS. model outperforms the base model, while the noEncQ. model performs slightly below the encS. model. This is expected, as the majority of questions in the Quality dataset do not contain PII information, making privacy preservation less relevant in most cases. To further investigate the effect of PII-specific questions, we conducted an additional experiment by using a subset of 35 questions that contain names (PII-Quality). The results show that in this subset, noEncQ. performs the same score as the base model, reinforcing our conclusion from the Judge Case dataset.

\subsection{Performance comparison}
\label{sec: experiment_perf_comp}
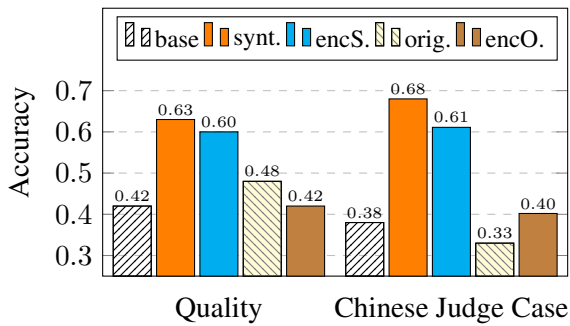
\begin{figure}[tbph]
\centering
\begin{tikzpicture}
\begin{axis}[ybar,
    bar width=0.5cm, 
    width=\columnwidth,
    height=\columnwidth/1.5,
    ylabel = {Accuracy},
    ymin=0.25, ymax=0.9,
    ytick={0.3,0.4,0.5,0.6,0.7},
    xmin=0.5, xmax=2.5,
    xtick = {1,2},
    xticklabels={
    Quality,
    Chinese Judge Case
    },
    major x tick style = transparent,
    legend pos=north west,
    legend columns=5,
    legend style={nodes={scale=0.855, transform shape}},
    ymajorgrids=true,
    grid style=dashed,
    nodes near coords,
    every node near coord/.append style={font=\tiny},
    point meta=rawy,
    every node near coord/.append style={inner xsep=1pt,yshift=-2pt, /pgf/number format/.cd,fixed zerofill,precision=2},
]

\addplot [color=black,postaction={pattern=north east lines}]  plot[ error bars/.cd, y dir=plus, y explicit] 
coordinates {(1,0.42) (2,0.38)};
\addlegendentry{base}
\addplot [color=black, fill=orange]  plot[ error bars/.cd, y dir=plus, y explicit] 
coordinates {(1,0.63) (2,0.68)};
\addlegendentry{synt.}
\addplot [color=black,fill=cyan]  plot[ error bars/.cd, y dir=plus, y explicit] 
coordinates {(1,0.60) (2,0.611)};
\addlegendentry{encS.}
\addplot [color=black,postaction={pattern=north west lines ,pattern color=gray},fill=yellow!20]  plot[ error bars/.cd, y dir=plus, y explicit] 
coordinates {(1,0.48) (2,0.33)};
\addlegendentry{orig.}
\addplot [color=black, fill=brown]  plot[ error bars/.cd, y dir=plus, y explicit] 
coordinates {(1,0.42) (2,0.402)};
\addlegendentry{encO.}
\end{axis}
\end{tikzpicture}
\caption{Performance comparison among base, trained models and crypto models.}
\label{fig: experiment_model_performance}
\end{figure}

To evaluate the effectiveness of our encryption-based approach, we conduct a performance comparison using five different models on the Quality dataset and Judge Case dataset. Table \ref{tab:experiment_performance_comp} shows the experimental data of model performance.
Figure \ref{fig: experiment_model_performance} shows the results for the Quality and Judge case datasets. In the Quality dataset, the synthetic model (synt.) significantly improves accuracy over the base model, achieving 0.63, while the encS. model reaches a comparable level. In contrast, continual pretraining on the original dataset (orig.) provides only a modest improvement (0.48), and encrypting the original dataset (encO.) eliminates this benefit, bringing accuracy back to the base level (0.42). The base model refers to the original model without pretraining on either dataset. The synthetic pretrained model (synt.) is pretrained on synthetic data, while the encrypted synthetic pretrained model (encS.) is pretrained on the encrypted version of the same synthetic data. The continually pretrained model on original data (orig.) is continually trained on original datasets. The continual pretrained model on encrypted original data (encO.) is trained on encrypted original datasets. The data for encS. model training is the encrypted version of synthetic data for the synt. model training.

On the Judge case dataset, where names and locations were encrypted in encS., we observe that while synthetic pretraining significantly boosts performance (0.68), encrypting PII leads to a slight drop to 0.61. In contrast, continual pretraining on the original data (orig.) and its encrypted counterpart (encO.) yields lower performance, reaching only 0.33 and 0.40, respectively. We further tested an encS. model trained on data where all PII (including bank card numbers, ID numbers, phone numbers, and dates, etc.) was encrypted, leading to a lower accuracy of 0.569. This suggests encrypting a wider range of PII impacts performance.

Overall, these results highlight that synthetic data effectively enhance model performance, while encryption allows for privacy preservation slightly degrading accuracy.

\subsection{Evaluation of RAG performance}
\label{sec: experiment_rag}
To evaluate retrieval-augmented generation (RAG) performance, we utilize word embeddings of the questions, generated using techniques from \citep{Chen2024m3} and \citep{2022allMiniLML6v2}. These embeddings are used to calculate the most relevant text by retrieving the top $K$ document chunks that are most contextually aligned with the question. After retrieval, the document chunks are reranked using methods proposed by \citep{Sun2024mair} and \citep{mxbai-rerank-large-v12024}, which helps refine the ranking to select the top $K$ results. The retrieved chunks are then used as a context for answering the question.
\begin{table}[bth]
    \centering
    \begin{tabular}{ccccccc}
    \hline
          Top-k & Retr. & Synt. & Encr. & Quality & Judge  \\
    \hline
          N/A & F & F & F & 0.424 & 0.388\\
          4 & T & F & F & 0.584 & \textbf{0.86}\\
          16 & T & F & F & \textbf{0.672} & 0.819\\
    \cline{2-6}
          N/A & F & T & F & 0.63 & 0.652 \\
          4 & T & T & F & \textbf{0.664} & \textbf{0.68}\\
          16 & T & T & F & 0.64 & 0.527\\
    \cline{2-6}
         N/A & F & T & T & 0.63 & 0.611\\
         4 & T & T & T & \textbf{0.664} & \textbf{0.847}\\
        16 & T & T & T & 0.456 & 0.736\\
    \hline
    \end{tabular}
    \caption{RAG performance comparison with the chunk size 1024. Retr., Synt. and Encr. denote for retrieval, synthesis and encryption, respectively.}
    \label{tab:experiment_rag}
\end{table}
Table \ref{tab:experiment_rag} displays the RAG performance results for both the Quality and Judge case datasets. We set the document chunk size to 1024 tokens and evaluated how the retrieved context influences the model performance. RAG consistently improves model performance across both datasets, with the greatest gains observed in models with partially encrypted data, while models with fully encrypted PII show less improvement - details are provided in Appendix~\ref{appendix:RAG}.

In the evaluation, we test the factors that attribute to model performance and validate the effectiveness and efficiency of our data synthesis method for pretraining. In Appendix \ref{appendix:hallucination}, we collect statistical data of hallucination in encrypted entities. An example of how the crypto model interacts with user queries is provided in Appendix \ref{appendix_demo}.
\section{Conclusion}
\label{sec:conclusion}
In this work, we explore the feasibility of continual pretraining of large language models using encrypted synthetic data generated through a weighted entity graph. By encrypting PII, our framework enables LLMs to learn from small, sensitive corpora while reducing the exposure of private information.
Across multiple experimental settings, models pretrained on encrypted synthetic data consistently outperform base models and achieve performance close to that of models trained on unencrypted synthetic data. We further demonstrate that encrypted models retain the ability to follow long-context instructions and effectively utilize retrieved information in retrieval-augmented generation scenarios. Our findings indicate that continual pretraining on encrypted data is a promising and practical direction for adapting LLMs in privacy-sensitive domains. We emphasize that this work represents an initial investigation into the design space, and that important consideration including the security trade-offs of deterministic encryption schemes and scalability to larger datasets remain open for future research.

\section*{Limitations}
Our approach has several limitations that point to important directions for future work.

\paragraph{Limitations of deterministic encryption.}
To enable consistent substitution of PII entities during continual pretraining, we adopt deterministic encryption. In our current implementation, this is realized using AES in ECB mode. While ECB preserves consistency across identical plaintexts, it is well known to be insecure for general-purpose data protection. In particular, ECB encrypts identical plaintext blocks into identical ciphertext blocks, thereby leaking equality patterns and frequency information. As a result, an adversary with access to the encrypted corpus may infer relationships between repeated entities or exploit statistical regularities, even without access to decryption keys \cite{ecb_insecurity_cryptoSE}. Moreover, ECB provides no semantic security and does not conceal structural patterns present in the underlying plaintext, making it unsuitable for strong privacy guarantees \cite{GOLDWASSER1984encryption}.

We emphasize that our use of ECB is motivated by functional requirements rather than cryptographic optimality, and our work does not claim formal security guarantees. Instead, it serves as an exploratory investigation into whether LLMs can be continually pretrained on encrypted representations while retaining utility. More robust deterministic encryption schemes, such as AES in SIV mode, offer stronger security properties while preserving determinism. AES-SIV is misuse-resistant and prevents many forms of information leakage associated with ECB, even under repeated encryption of identical plaintexts \cite{cryptoeprint:2006/221}. Their impact on training efficiency and model performance remains an open question. We leave a systematic evaluation of stronger deterministic encryption alternatives to future work.

\paragraph{Impact of encryption on representation and context length.}
The ciphertext used to replace PII entities lacks semantic meaning and typically spans more than ten tokens per entity. This expansion can increase sequence length and may impair the model’s ability to handle long contexts effectively. Furthermore, we do not explicitly analyze how encrypted entities alter sentence-level semantics coherence, leaving the representational effects of encryption on downstream reasoning underexplored.

\paragraph{Dataset scale and generality.}
Our method is designed and evaluated on small, domain-specific corpora, where manual verification of entity extraction and encryption is feasible. Standard large-scale public benchmarks are not directly applicable due to the encrypted nature of the training data, and the scalability of our approach to substantially larger datasets remains uncertain. Evaluating encrypted pretraining at scale and across broader domains is an important direction for future work.

\paragraph{Reliance on LLM-based synthesis.}
Finally, since large language models are used for both entity extraction and data synthesis, there is an inherent risk of generating fabricated or hallucinated content. Eliminating such artifacts entirely remains challenging and may affect the quality of the synthesized data.

\section*{Ethical Considerations}
\label{sec:limitation}
Our work is developed with the primary goal of enhancing privacy when large language models continually pretrain on private data. Rather than introducing new ethical risks, our framework seeks to mitigate privacy concerns by ensuring that models can learn from encrypted synthetic data without accessing or memorizing real personally identifiable information. Our encryption before synthesis method with minimal manual verification could guarantee that there is no privacy leakage. We do not use unauthorized data or release any private or sensitive datasets in this work. We mask personal data in the example of Chinese judge case QA questions. PII is encrypted in the original data and synthetic data. Furthermore, the synthetic samples themselves are not released to the public or any third party. We encourage future work to explore fairness, bias mitigation, and robustness in encrypted continual pretraining regimes.

\section*{Information about Use of AI Assistants}
We use openAI chatGPT as an assistance purely with the language of the paper.

\bibliography{custom}

\appendix
\label{sec:appendix}

\section{Additional Experimental Details and Results}
We provide several additional experiments and training hyperparameters for the main results. In additional experiments, first, we test whether the ciphertext generated by crypto LLMs is able to decode to correct entities and classify the type of ciphertext that fails to decode the corresponding entity. Then, we evaluate the hyperparameters tuning for RAG (chunk size and rerank top-k). The percentage of failure responses that do not follow the format and prompt instructions in RAG is evaluated. We provide the training parameters. Additionally, we evaluate the privacy leakage when the data synthesis is performed before PII encryption.

\subsection{Hallucination on Ciphertext}
\label{appendix:hallucination}
We decrypt all ciphertext in LLM responses article by article on the Judge case dataset. There are two main types of failure cases. The crypto model cites ciphertext from other articles to answer the multiple choice questions, although the ciphertext is able to decrypted. In the other case, the model generates ciphertext that is not exist in the synthesized data, and it is not able to be decrypted to meaningful plaintext by the private keys.

Table \ref{tab:appendix_hallucination} depicts the statistical results of the ciphertext on the Judge case dataset. There are 68 unique cipher entities generated in responses where person's name and location are encrypted, and 135 unique cipher entities where all PII entities are encrypted. The percentage of hallucination of those two crypto models are close around 15\%. There are only 5 cipher entities that do not exist in the synthesized corpus, the other hallucination is attribute to cite encrypted text from other articles. 
\begin{table}[htbp]
    \centering
    \begin{tabular}{cccc}
    \hline
        &\#unique failure & \#unique cipher & ratio \\
    \hline
       PL & 11 & 68 &  0.16\\
       ALL & 23 & 135 &  0.17\\
    \hline
        & total \#failure & total \#cipher & ratio \\
    \hline
        PL &  36 & 1236 & 0.029\\
       ALL &  72 & 1850 & 0.039 \\
    \hline
    \end{tabular}
    \caption{Statistical results on ciphertext hallucination.}
    \label{tab:appendix_hallucination}
\end{table}
There are 1000 to 2000 times of ciphertext in the response, and less than 5\% of cipher entities are fabricated content. 

We analyze exceptional ciphertexts across two datasets - Chinese Judge and English Quality - using Qwen2.5 7B and LLaMA3 8B models. We categorize failures into two types: (1) FCAOA (Failures Caused by Ciphertexts Appearing in Other Articles): The model erroneously outputs ciphertexts copied from unrelated contexts; (2) FCND ((Failures Caused by Non-Decodable Ciphertexts): The output ciphertexts are malformed and cannot be decoded, and they do not match any known synthetic data.

There are two primary technical issues that lead to decryption failures:
\begin{itemize}
    \item Base64 Format Issues: The encryption pipeline uses Base64 encoding to represent encrypted bytes. If the resulting string is not a multiple of 4 characters, Base64 decoding fails.
    \item PKCS7 Padding Errors: If the original plaintext was not padded to a multiple of 16 bytes, decryption will raise an "invalid padding bytes" error.
\end{itemize}
In the Chinese Judge dataset, 5 entities failed to decrypt: 2 due to Base64 formatting errors and 3 due to PKCS7 padding errors. In the Quality dataset, all failures were attributed to FCAOA (i.e., ciphertexts copied from other articles), see Table \ref{tab:decrypt_fail_cases}.
\begin{table}[hptb]
    \centering
    \begin{tabular}{cccc}
    \hline
        Model & \#Fialures & \#FCAOA & \#FCND \\
    \hline
        Qwen2.5 7B & 72 & 67 & 5 \\
        Llama3 8B & 3 & 3 & 0 \\
    \hline
    \end{tabular}
    \caption{Decryption failure statistics by model and failure type.}
    \label{tab:decrypt_fail_cases}
\end{table}

To address exceptional ciphertexts, we introduce two fallback mechanisms. When Base64 decoding fails due to incorrect padding, we attempt to recover by appending = characters until the string length is a multiple of 4. If decoding succeeds, the output is flagged as a revised Base64 string. In the PKCS7 padding error, when decryption fails in the unpadding step, we mark the output with a padding error flag. Due to uncertainty about the root cause, we do not apply automatic correction and instead mark these cases for manual review.

\subsection{RAG results and hyperparameters effect}
\label{appendix:RAG}
We compare the RAG performance with the fixed chunk size - 1024. We set the rerank top-k $\in \{4,16\}$ in the main experiments. Table \ref{tab:experiment_rag} shows the statistical results. In the Quality dataset, RAG leads to performance improvements across all models. The base model sees a significant increase in accuracy from 0.424 to 0.672, while the synt. and encS. models experienced slight increases from 0.63 to 0.664. However, with RAG, all models converged to a similar performance level. In the Judge Case dataset, the base model and the encS. model (which encrypts only names and locations) increased significantly to a similar level (from 0.611 to 0.847), while models trained on data with fully encrypted PII exhibit less improvement (from 0.569 to 0.625). This performance gap is attributed to the pretraining process that affects the long instruction-following capabilities. Many continual pretraining models failed to generate responses in the expected format, deviating from the answer structure provided in the prompt.

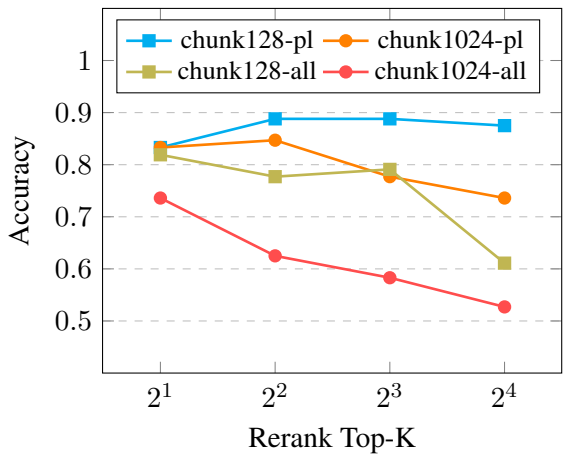
\begin{figure}[ht]
\centering
\begin{tikzpicture}
\begin{axis}[
    xlabel={Rerank Top-K},
    ylabel={Accuracy},
    width=\columnwidth*0.99,
    height=\columnwidth/1.2,
    ymin=0.4, ymax=1.1,
    ytick={0.5,0.6,0.7,0.8,0.9,1.0},
    xmin=1.41, xmax=22.6,
    xtick={2,4,8,16},
    legend pos=north west,
    legend style={nodes={scale=0.9, transform shape}, legend columns=2},
    ymajorgrids=true,
    grid style=dashed,
    xmode=log, 
    log basis x=2, 
]
\addplot [color=cyan,mark=square*, line width=1pt,mark size=2pt] coordinates {
    (2,0.833) (4,0.888) (8,0.888) (16,0.875)
};
\addlegendentry{chunk128-pl}
\addplot [color=orange,mark=*, line width=1pt,mark size=2pt] coordinates {
    (2,0.833) (4,0.847) (8,0.777) (16,0.736)
};
\addlegendentry{chunk1024-pl}
\addplot [color=yellow!70!black, mark=square*,line width=1pt,mark size=2pt] coordinates {
  (2,0.819) (4,0.777) (8,0.791) (16,0.611)
};
\addlegendentry{chunk128-all}
\addplot [color=red!70,mark=*,line width=1pt,mark size=2pt] coordinates {
  (2,0.736) (4,0.625) (8,0.583) (16,0.527)
};
\addlegendentry{chunk1024-all}
\end{axis}
\end{tikzpicture}
\caption{The effect of RAG hyperparaters on accuracy.}
\label{fig: appendix_rag_hyper}
\end{figure}

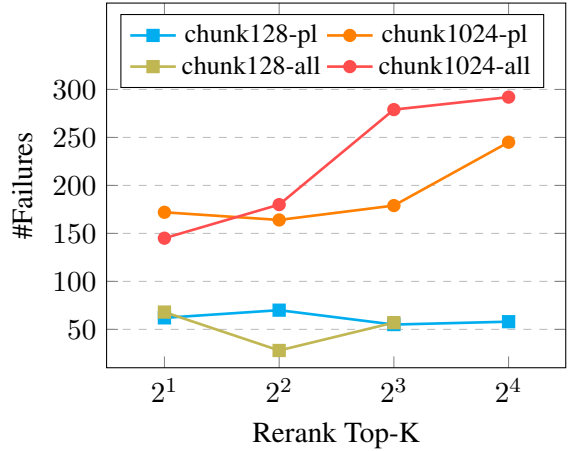
\begin{figure}[ht]
\centering
\begin{tikzpicture}
\begin{axis}[
    xlabel={Rerank Top-K},
    ylabel={\#Failures},
    width=\columnwidth*0.99,
    height=\columnwidth/1.2,
    ymin=10, ymax=390,
    ytick={50,100,150,200,250,300},
    xmin=1.41, xmax=22.6,
    xtick={2,4,8,16},
    legend pos=north west,
    legend style={nodes={scale=0.9, transform shape}, legend columns=2},
    ymajorgrids=true,
    grid style=dashed,
    xmode=log, 
    log basis x=2, 
]
\addplot [color=cyan,mark=square*, line width=1pt,mark size=2pt] coordinates {
    (2,62) (4,70) (8,55) (16,58)
};
\addlegendentry{chunk128-pl}
\addplot [color=orange,mark=*, line width=1pt,mark size=2pt] coordinates {
    (2,172) (4,164) (8,179) (16,245)
};
\addlegendentry{chunk1024-pl}
\addplot [color=yellow!70!black, mark=square*,line width=1pt,mark size=2pt] coordinates {
  (2,68) (4,28) (8,57)
};
\addlegendentry{chunk128-all}
\addplot [color=red!70,mark=*,line width=1pt,mark size=2pt] coordinates {
  (2,145) (4,180) (8,279) (16,292)
};
\addlegendentry{chunk1024-all}
\end{axis}
\end{tikzpicture}
\caption{The effect of RAG hyperparaters on instruction following.}
\label{fig: appendix_rag_hyper_failure}
\end{figure}

Here, we evaluate the performance of RAG results in accuracy and long-instruction following capability varying hyperparameters. Figure \ref{fig: appendix_rag_hyper} shows the RAG accuracy by varying the rerank top-k and chunk size. The line chart \ref{fig: appendix_rag_hyper_failure} illustrates the number of LLM responses that do not follow the instruction to provide answers with different hyperparameters. The performance with a small chunk size is better than that with long chunk size. As the length of context increases, the accuracy gradually decreases. There are more failure cases with large chunk sizes. When crypto. LLM takes more samples of chunks as contexts, the responses tend to break the instructions. Those indicates that the continual pretrained model hamper the original capability on the long-context instruction following. Taking small retrieved chunks as contexts will maximize RAG performances. \textit{chunk128-pl} denotes that the chunk size is 128 and the personal names and locations are encrypted. \textit{chunk128-all} denotes that the chunk size is 128 with all encrypted PII. 

\subsection{Training hyperparameters}
\label{appendix:train_parameter}
Table \ref{tab:train_hyperparam} shows the hyperparameters of continual pretraining of our main experiments.
\begin{table}[hbtp]
    \centering
    \begin{tabular}{cc}
    \hline
        \textbf{Name} & \textbf{Value} \\
    \hline
        Learning rate & 5e-6 \\
        Training epochs & 2 \\
        Batch size per device & 2 \\
        Gradient accumulation steps & 4\\
        lr scheduler type & cosine \\
        warmup ratio & 0.1  \\
        cutoff length & 2048 \\
    \hline
    \end{tabular}
    \caption{Training hyperparameters.}
    \label{tab:train_hyperparam}
\end{table}



\subsection{Privacy leakage in data Synthesis before PII encryption}
\label{appendix_privacy_leakage}

To quantify potential privacy leakage in synthetic data generated before applying PII encryption, we conducted both empirical and manual evaluations on the Chinese Judge dataset. Encrypted entities were identified by matching structured ciphertext patterns (e.g., \texttt{Person\_[cipher]} for names). For unencrypted entities, we manually extracted all PII from the original articles and searched for their occurrences in the synthetic dataset.

Table~\ref{tab:privacy_leakage} summarizes the total number of encrypted and unencrypted entities across all synthetic data. While encrypted entities number in the millions, unencrypted instances are relatively rare. The overall ratio of unencrypted to encrypted entities is approximately 1:300, suggesting a low risk of privacy leakage. We assume zero leakage for entities such as ID numbers, phone numbers, bank card numbers, and dates, as they are consistently identified using regular expressions and encrypted accordingly.

\begin{table}[hptb]
    \centering
    \begin{tabular}{cccc}
    \hline
        Entity type  & \#Unenc.  & \#Enc. & ratio\\
    \hline
        Name & ~3K & ~1M & 3:1000 \\
        Location & 947 & ~68K & 12:1000 \\
        Others & 0 & ~195K & N/A \\
        Total & ~4K & ~1.2M & 1:300\\
        \hline
    \end{tabular}
    \caption{Ratio between unencrypted and encrypted entities in entire synthetic data. \#Unenc. and \#Enc. denote for number of unencrypted entities and number of encrypted entities.}
    \label{tab:privacy_leakage}
\end{table}

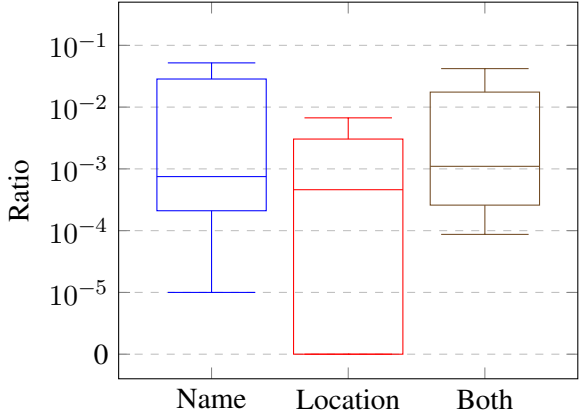
\begin{figure}
\centering
\begin{tikzpicture}
\begin{axis}[
    boxplot/draw direction=y,
    ylabel={Ratio},
    xtick={1,2,3},
    xticklabels={Name, Location, Both},
    ytick={0.1,0.01,0.001,0.0001,0.00001,0.000001},
    yticklabels={$10^{-1}$,$10^{-2}$,$10^{-3}$,$10^{-4}$,$10^{-5}$, 0},
    ymin=0.0000004, ymax=0.5,
    ymajorgrids=true,
    grid style=dashed,
    ymode=log,              
    log basis y=10,         
    width=\columnwidth*0.99,
]
\addplot+[
    boxplot prepared={
        median=0.00075,
        upper quartile=0.0285,
        lower quartile=0.00021,
        upper whisker=0.052,
        lower whisker=0.00001
    },
    color=blue,
    boxplot prepared
] coordinates {};
\addplot+[
    boxplot prepared={
        median=0.00046,
        upper quartile=0.00304,
        lower quartile=0.000001,
        upper whisker=0.0067,
        lower whisker=0.000001
    },
    color=red,
    boxplot prepared
] coordinates {};
\addplot+[
    boxplot prepared={
        median=0.0011,
        upper quartile=0.0175,
        lower quartile=0.000259,
        upper whisker=0.042,
        lower whisker=0.000087
    },
    boxplot prepared
] coordinates {};
\end{axis}
\end{tikzpicture}
\caption{Box plot of ratio between unencrypted and encrypted entities.}
\label{fig:box_plot_ratio}
\end{figure}
To analyze variance and consistency across different samples, we further collected this ratio across 7 source articles, each with 5 synthetic outputs, yielding a total of 35 samples. We present the distribution of leakage ratios grouped by entity type. Figure~\ref{fig:box_plot_ratio} presents a box plot of the ratio between unencrypted and encrypted entities for three categories: names, locations, and the combined set of both. For names, the median leakage ratio is around $7.5 \times 10^{-4}$, with some outliers reaching as high as 0.05. Location entities exhibit a slightly lower median of $4.6 \times 10^{-4}$. The combined group shows a slightly higher median but similarly sparse lower outliers. These results indicate that while privacy leakage is generally low across most samples, certain cases do exhibit relatively higher exposure. However, our approach by applying robust PII encryption before data synthesis effectively mitigates the risk of private data leakage, even in rare outlier cases.

\section{Chinese Judge Case Dataset and QA Examples.}
\label{appendix_chinese_dataset}
The Chinese Judge case dataset consists of multiple civil litigation cases, focusing mainly on legal proceedings. Each case includes detailed information such as the case ID, court details, plaintiff and defendant private information, and legal arguments from both sides. We provide questions and answers based on various case details including the responsibilities of the parties involved and their personal information. These questions are designed to test the model's ability to process sensitive legal information, understand the relationships between entities, and assess the accuracy of legal reasoning. The dataset is useful for training and evaluating models on handling sensitive legal data while ensuring privacy, as it involves detailed personal and financial information. 

\begin{table}[hptb]
    \centering
    \begin{tabular}{p{0.9\columnwidth}}
    \hline
    \rule{0pt}{11pt}
         \textbf{Question:} \begin{CJK*}{UTF8}{gbsn} 宋**与刘**的离婚协议中约定: 婚生子宋**由谁抚养？ \end{CJK*}  \\
         \ \textbf{Translation: } In the divorce agreement between Song ** and Liu **, who is designated to have custody of their kid, Song **? \\
         \ \ \ A. \begin{CJK*}{UTF8}{gbsn} 陈**抚养 \end{CJK*} \hspace{0.86cm} (Chen ** has custody)  \\
         \ \ \ B. \begin{CJK*}{UTF8}{gbsn} 宋**抚养 \end{CJK*} \hspace{0.86cm} (Song ** has custody)\\
         \ \ \ C. \begin{CJK*}{UTF8}{gbsn} 双方共同抚养  \end{CJK*} \ \ (Joint custody)\\
         \ \ \ D. \begin{CJK*}{UTF8}{gbsn} 刘**抚养 \end{CJK*} \hspace{0.86cm} (Liu ** has custody)\\
    \hline
    \end{tabular}
    \caption{a QA example in  the Chinese Judge dataset (sensitive contents are masked).}
    \label{tab:appendix_qa_example}
\end{table}
In addition, Table \ref{tab:appendix_qa_example} presents a representative QA example from the Chinese Judge dataset, where sensitive information has been masked for privacy protection. This example collectively supports the evaluation of reasoning capabilities in crypto. language models.

\section{Entity Association Strength Prompt}
\label{appendix_prompt}
The entity association strength prompt (Table \ref{tab:appendix_asso_strength}) was designed to guide the model in evaluating nuanced relationships between two entities within a legal or factual context. It provides structured relationship, dependency, and indirect association to ensure consistent and explainable scoring across cases. This prompt enables fine-grained understanding of inter-entity dynamics, which is particularly useful in tasks such as knowledge extraction or internal relation analysis.
\begin{figure*}[hptb]
\begin{tcolorbox}[enhanced,
  colback=gray!10,       
  colframe=black!70,     
  coltitle=white,        
  fonttitle=\bfseries,   
  colbacktitle=black!80, 
  rounded corners,
  boxrule=0.8pt,
  title={Prompt}]

         You will act as a scorer responsible for evaluating the strength of association between two specific entities (E1 and E2) in a given article. The evaluation will be based on the following three criteria:

        1. Causal Relationship: Assess whether there is a causal relationship between E1 and E2. Describe each of the following situations:
        
           - E1 causes or influences E2: <specific description>
           
           - E2 causes or influences E1: <specific description>
    
    2. Dependency: Evaluate the extent to which one entity depends on or is influenced by the other within the context of the article.
    
       - E1 depends on E2: <specific description>
       
       - E2 depends on E1: <specific description>
    
    3. Indirect Association: If there is no direct relationship between E1 and E2, identify indirect connections through other entities.
    
       * List intermediary entities mentioned in the article that are connected to both E1 and E2.
       
       * Describe how these intermediary entities help establish a connection between E1 and E2.
       
       * Determine the type of relationship between the intermediary entities and each of the two target entities.
    
    After analyzing these three criteria, assign a score (between 0 and 1) representing the overall strength of association between E1 and E2 in the context of the article. The score should reflect the degree of connection based on the article’s content.\\
    \\
    Your response should follow the format below to ensure clarity and organization:
    
    \#\#\# Causal Relationship between <E1> and <E2> in <Title>
    
    E1 influences E2: <Describe the causal effect of E1 on E2>.
    
    E2 influences E1: <Describe the causal effect of E2 on E1>.
    
    \#\#\# Dependency between <E1> and <E2> in <Title>
    
    E1 depends on E2: <Describe how E1 depends on E2>.
    
    E2 depends on E1: <Describe how E2 depends on E1>.
    
    \#\#\# Indirect Association between <E1> and <E2> in <Title>
    
    Intermediary Entities: <List and describe other entities that act as bridges between E1 and E2>.
    
    Link through Intermediaries: <Discuss how these entities link E1 and E2>.
    
    Score: <Assign a score between 0 and 1 based on the strength of connection>. 
    \end{tcolorbox}
    \caption{Entity association strength prompt.}
    \label{tab:appendix_asso_strength}
\end{figure*}
\section{Example Appendix}
\label{appendix_demo}
\begin{figure*}
    \centering
    \includegraphics[width=1\linewidth]{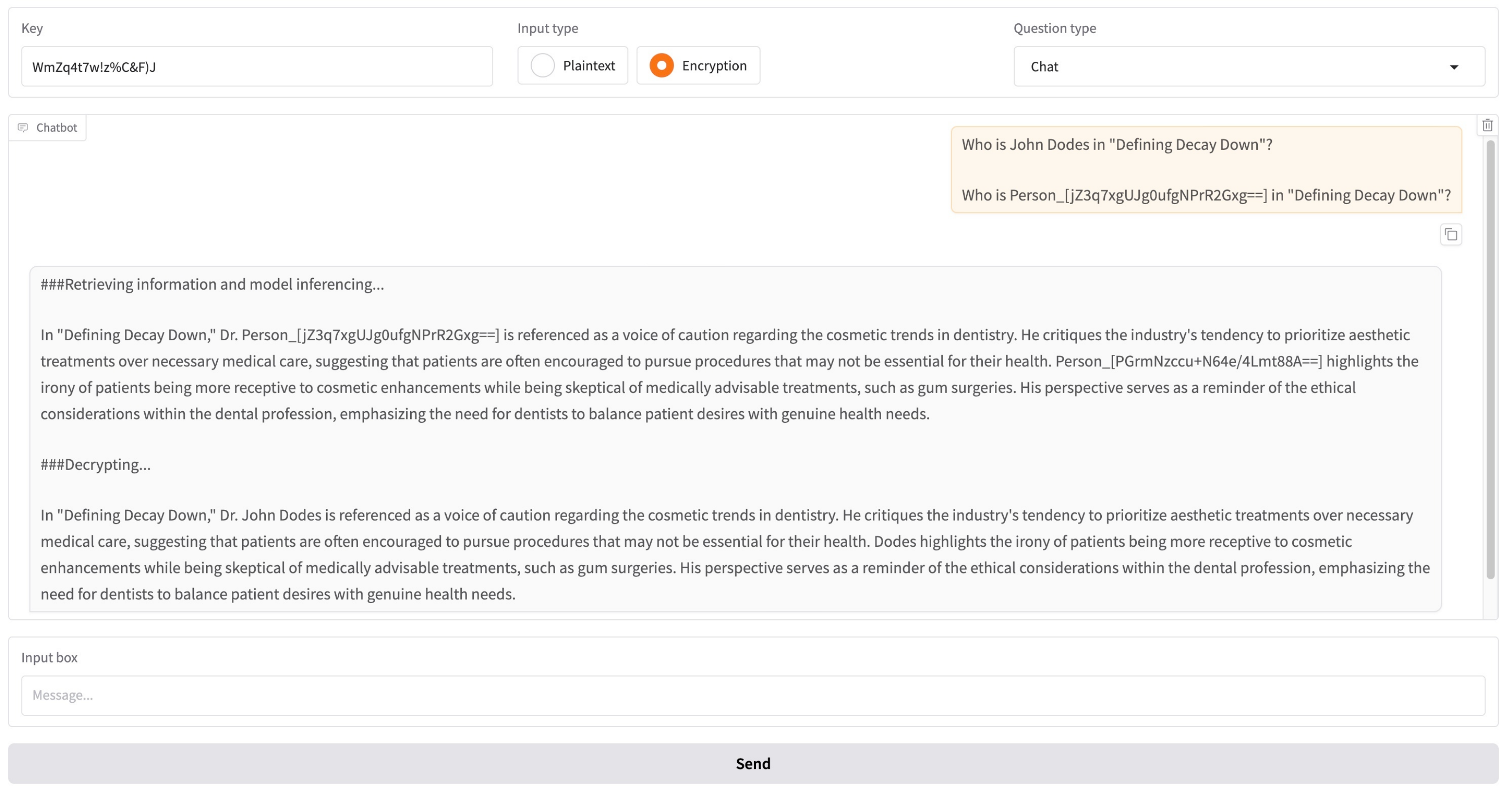}
    \caption{A demonstration on the QuALITY dataset with encrypted personal names.}
    \label{fig:appendix_demo}
\end{figure*}

In this section, we provide a demonstration of the interaction between the user and the crypto model. The crypto model is continually pretrained on synthesized cipher data of QuALITY with encrypted personal names. Figure \ref{fig:appendix_demo} demonstrates the process of model response using RAG to retrieve chunks from encrypted original articles. We use the all-MiniLM-L6-v2 and the mxbai-rerank-large-v1 for retrieval and reranking, respectively \cite{2022allMiniLML6v2, mxbai-rerank-large-v12024}. The user input will be encrypted by a key. Then, models will retrieve and rerank 4 chunks as the contexts. The crypto model will respond to the user in ciphertext with the contexts as prompts. Finally, the user will decrypt the response to get the result.

\end{document}